\documentclass[a4paper,10.5pt,twoside]{article}

\usepackage[english]{babel}
\usepackage{graphicx}
\usepackage{verbatim}
\usepackage{amsfonts}
\usepackage{makeidx}
\usepackage{color}
\usepackage{amsmath}
\usepackage{epstopdf}
\usepackage{hyperref}
\usepackage{cite}
\usepackage{wrapfig}
\usepackage{hyperref}
\usepackage{setspace}

  \renewenvironment{thebibliography}[1]{%
    \begin{oldthebibliography}{#1}%
      \setlength{\parskip}{0ex}%
      \setlength{\itemsep}{1ex}%
  }%
  {%
    \end{oldthebibliography}%
  }

\newcommand{\pp}{\mbox{pp}}

\newcommand{\GeV}{\mathrm{GeV}}

\newcommand{\PbPb}{\mbox{Pb--Pb}}

\newcommand{\pt}{p_{\rm t}}

\newcommand{\DtoKpi}{{\rm D^0\to K^-\pi^+}}
\newcommand{\DtoKpipi}{{\rm D^+\to K^-\pi^+\pi^+}}
\newcommand{\DstartoDpi}{{\rm D^{*+}\to D^0\pi^+}}
\newcommand{\Dzero}{{\rm D^0}}
\newcommand{\Dstar}{{\rm D^{*+}}}
\newcommand{\Dplus}{{\rm D^+}}

\title{Heavy Flavour Measurements in $\pp$ and $\PbPb$ Collisions with the ALICE Experiment at LHC}
\author{C. Zampolli{\footnote{INFN Sezione di Bologna, \href{mailto:Chiara.Zampolli@cern.ch}{Chiara.Zampolli@cern.ch}}} \, for the ALICE Collaboration} 

\begin{document}
\maketitle

\begin{abstract}
Heavy flavour is mainly produced during the initial hard partonic interactions in a heavy ion collision, and is well-suited to probe the early phases of the evolution of the system. 
This contribution will focus on  $\PbPb$ analyses at a centre-of-mass energy per nucleon pair of 2.76 TeV, with some hints at the $\pp$ data at 7 and 2.76 TeV. Results of open heavy flavour analyses  will be shown for various decay channels, including electrons, muons, and hadronic charm decays, together with results obtained for heavy quarkonia at both central and forward rapidities.
\end{abstract}

\section{Introduction}
The extreme energy density 
and temperature conditions reached in $\PbPb$ collisions at LHC energies
are expected to 
generate a deconfined plasma of quarks and gluons (the so-called Quark-Gluon Plasma, QGP~\cite{Karsch}), from which a phase transition to ordinary colourless hadronic matter takes place
as a consequence of subsequent expansion and cooling down. Among the various observables that shed light on the properties
of the QGP, heavy flavour production is expected to provide unique and important probes. Since heavy quarks (charm and beauty) are
produced during the early stages of the collisions, they experience the whole 
evolution of the system. High momentum open charm and beauty are of great interest for the role they play in the 
study of the in-medium partonic energy loss. Such energy loss occurs due to elastic processes (collisional energy loss)~\cite{thoma}, as well as inelastic processes~\cite{gyulassy,bdmps} that are expected to dominate at high momentum. Due to the smaller QCD colour coupling (Casimir factor) for quarks than for gluons, the energy loss expected for quarks should be smaller than that for gluons. Since light flavour hadrons arise dominantly form gluon jets, this effect may exhibit itself in different suppression patterns for light and heavy flavour hadrons. Moreover the ``dead cone effect" should reduce the radiation at small angles for heavy quarks, imposing a further hierarchy on the energy loss depending on the mass
of the quark~\cite{EnergyLoss}.

The study of quarkonium can provide important insight on the properties of the medium. Production of
charmonium  in a QGP is determined by the mechanism of colour screening~\cite{Matsui}, which
leads to a suppressed production in a hot medium compared to $\pp$ collisions. Moreover, the abundant production of charm quark pairs
at LHC energies could lead to a new mechanism of charmonium formation via recombination of uncorrelated charm-anticharm quarks in a deconfined 
medium~\cite{BraunMunzinger:2000px,Thews:2000rj}, which could even result (depending on the overall charm abundance) in a J/$\psi$ production in $\PbPb$ enhanced with respect to $\pp$ 
collisions.



Heavy quarks at the LHC are an important probe also in $\pp$ collisions. Such measurements enable tests of theoretical calculations based on perturbative QCD 
for heavy quark production in a new
energy regime. They could shed light on the hadroproduction mechanisms that govern quarkonia production. This is an important outstanding problem, since current theoretical models cannot satisfactorily describe measurements of heavy quark rapidity and $\pt$ distributions, and  polarization. $\pp$ data also provide an essential baseline for $\PbPb$ measurements, the 
comparison with which is needed to
discriminate between initial and final effects in their production.

\begin{figure}[t!]
\begin{center}
\includegraphics[height=6cm]{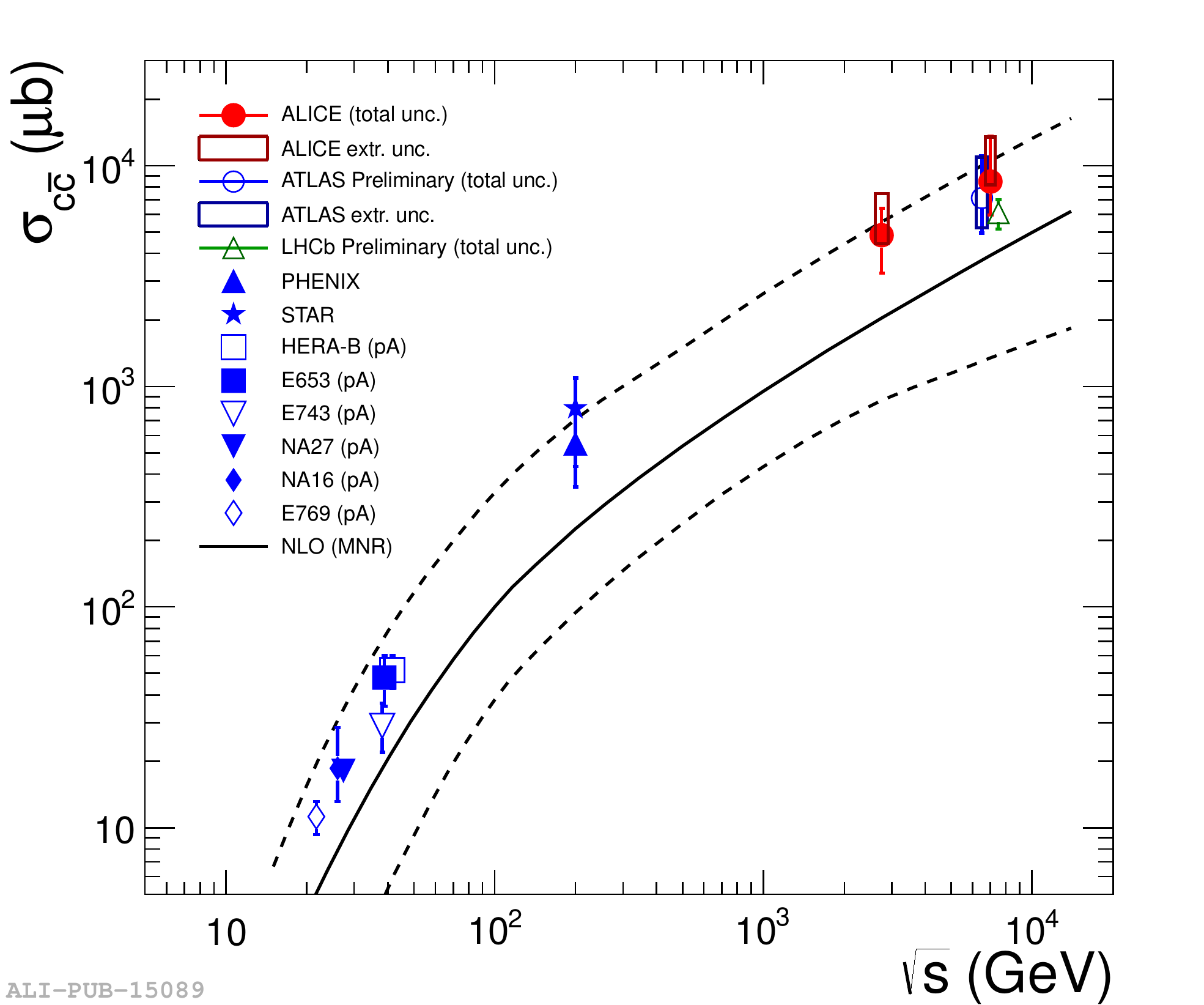} 
\end{center}
\vspace{-0.5cm}
\caption{Charm-anticharm cross section for the full $\pt$ and rapidity phase space 
as a function of the centre-of-mass energy for different experiments at HERA, FNAL, SPS, RHIC and the LHC, for different collisions systems ($\pp$ collisions for the LHC points)~\cite{crossSection:2012sx,LHCbpoint,ATLASpoint,Lourenco:2006vw,STAR:2011aa,Adare:2010de}. The LHCb and ATLAS measurements are slightly shifted for visibility. The NLO MNR calculation is also 
drawn~\cite{Mangano:1991jk}.}
\label{figure:HFProduction}
\end{figure}

Figure~\ref{figure:HFProduction} shows the total charm-anticharm cross section in the full phase space as a function of the centre-of-mass for different experiments (~\cite{crossSection:2012sx} and 
references therein). 
The ALICE values refer to the measurements at 
$\sqrt{s} = 2.76$ and 7 TeV and were extrapolated down to $\pt$ = 0 and at full rapidity using FONLL calculations. The LHCb~\cite{LHCbpoint} and ATLAS~\cite{ATLASpoint} points are drawn as well. As one can see, the values at the LHC are $\sim 10$ times larger than those at RHIC, making the LHC a ``heavy flavour factory".

The paper is organized as follows. First, a brief description of the ALICE experiment will be given in Section~\ref{sec:ALICE} focusing on the detectors involved in the heavy flavour studies.
In Sec.~\ref{sec:D2H} the results for the open heavy flavour analyses will be presented. These include the open charm mesons in the decay channels $\DtoKpi$, $\DtoKpipi$ and $\DstartoDpi$, 
and the analysis of the semi-electronic decays
$\rm{D},\rm{B} \rightarrow \rm{e} + \rm{X}$ at midrapidity (more details can be found in~\cite{RossiHere}), together with the study of the heavy flavour muons at forward rapidity (see also ~\cite{Loic}). The 
results for the $\rm{J}/\psi$ analysis in the electronic and muonic decay channel, at central and forward rapidity respectively, will be presented in Sec.~\ref{sec:Jpsi} (for more details see also~\cite{Carmelo}). 
Finally, some concluding remarks will be given in 
Sec.~\ref{sec:conclusions}.

\section{The ALICE experiment and its heavy flavour program}
\label{sec:ALICE}
ALICE~\cite{PPR1} is the experiment at the Large Hadron Collider (LHC) devoted to the study of ultra-relativistic heavy-ion collisions. The ALICE heavy flavour program relies on the excellent tracking and particle identification capabilities of the experiment. In the ALICE central barrel, covering the 
pseudorapidity range $|\eta| < 0.9$, in a 
magnetic field of 0.5 T, the Inner Tracking System (ITS) and the Time Projection Chamber (TPC) are the main tracking devices. 
The ITS is primarily dedicated to the reconstruction of the primary and secondary vertices, with a transverse impact parameter resolution better than 75 $\mu$m
for $\pt > 1$ $\GeV/c$. The two innermost layers of the ITS, the Silicon Pixel Detector (SPD), are also used to provide the minimum-bias trigger
to the experiment together with the VZERO detector\footnote{The ALICE VZERO detector, covering the pseudorapidity regions $2.8 < \eta < 5.1$ and $-3.7 < \eta < -1.7$, is made up of two arrays of scintillator tiles on both sides of the interaction point, and is aimed at providing the minimum-bias trigger, and centrality and luminosity information.}. 
Besides tracking and momentum determination, the TPC is used in the heavy flavour analyses also for particle identification (PID) of charged particles via specific energy deposition $dE/dx$. 
Charged hadrons at intermediate momenta are identified by the ALICE Time Of Flight (TOF) detector. Electron PID is carried out for $\pt$ $< 4$ GeV$/c$ by the TPC and the TOF detectors, while in the range $\pt > 1$ GeV$/c$ the Transition Radiation Detector (TRD) is used. The Electromagnetic Calorimeter (EMCAL) 
identifies electrons at high momenta ($\pt > 5$ GeV$/c$). Finally, the ALICE Muon Spectrometer at $-4 < \eta < -2.5$ performs both the reconstruction and the identification of muon tracks with
$\pt > 4$ GeV$/c$. 

\section{Open heavy flavour}
\label{sec:D2H}
In this section, the main ALICE open heavy flavour analyses will be discussed. The results included here will cover both data collected in $\pp$ collisions at $\sqrt{s} = 7$ TeV and $\PbPb$ collisions
at $\sqrt{s_{NN}}$ = 2.76 TeV. Emphasis will be put on the nuclear modification factor $R_{AA}$ defined as:
\begin{equation}
\label{eq:RAA}
R_{AA}(\pt) = \frac{d^2N^{AA}/d\eta d \pt}{\langle T_{AA}\rangle d^2\sigma^{pp}/d\eta d \pt}
\end{equation}
Here, $\langle T_{AA}\rangle$ is the average nuclear overlap function for a given centrality class and is proportional to $\langle N_{coll}\rangle$. 
$R_{AA}$ quantifies the effect  
of the hot and dense medium created in heavy-ion collisions by comparing the production yield in $\pp$ collisions, after an appropriate scaling via the 
number of binary collisions, to that obtained in $\PbPb$ interactions. 
Any observed deviation of $R_{AA}$ from unity is due to effects specific to nuclear collisions. 

\subsection{D mesons}
\label{sec:openCharm}
The hadronic decay channels $\DtoKpi$, $\DtoKpipi$, and $\DstartoDpi$ were used for the study of open charm production. The analysis of these channels
is based on the topology of the 
decay, characterized by a secondary vertex displaced from the primary one, and exploits the excellent vertex reconstruction and tracking capabilities of the ALICE ITS and TPC detectors.
The background is reduced at low transverse
momenta by combining the TPC and TOF PID information. In order to minimize as much as possible the signal loss due to the 
usage of the PID, the rejected candidates are those for which the daughters have been identified by both TPC and TOF, but the two detectors provide different PID responses. In addition to such cases, 
also candidates for which the two daughter tracks were both identified as pions or as kaons were discarded.

\begin{figure}[t!]
\begin{center}
\begin{tabular}{ccc}
\hspace{-1.cm}
\includegraphics[height=5cm]{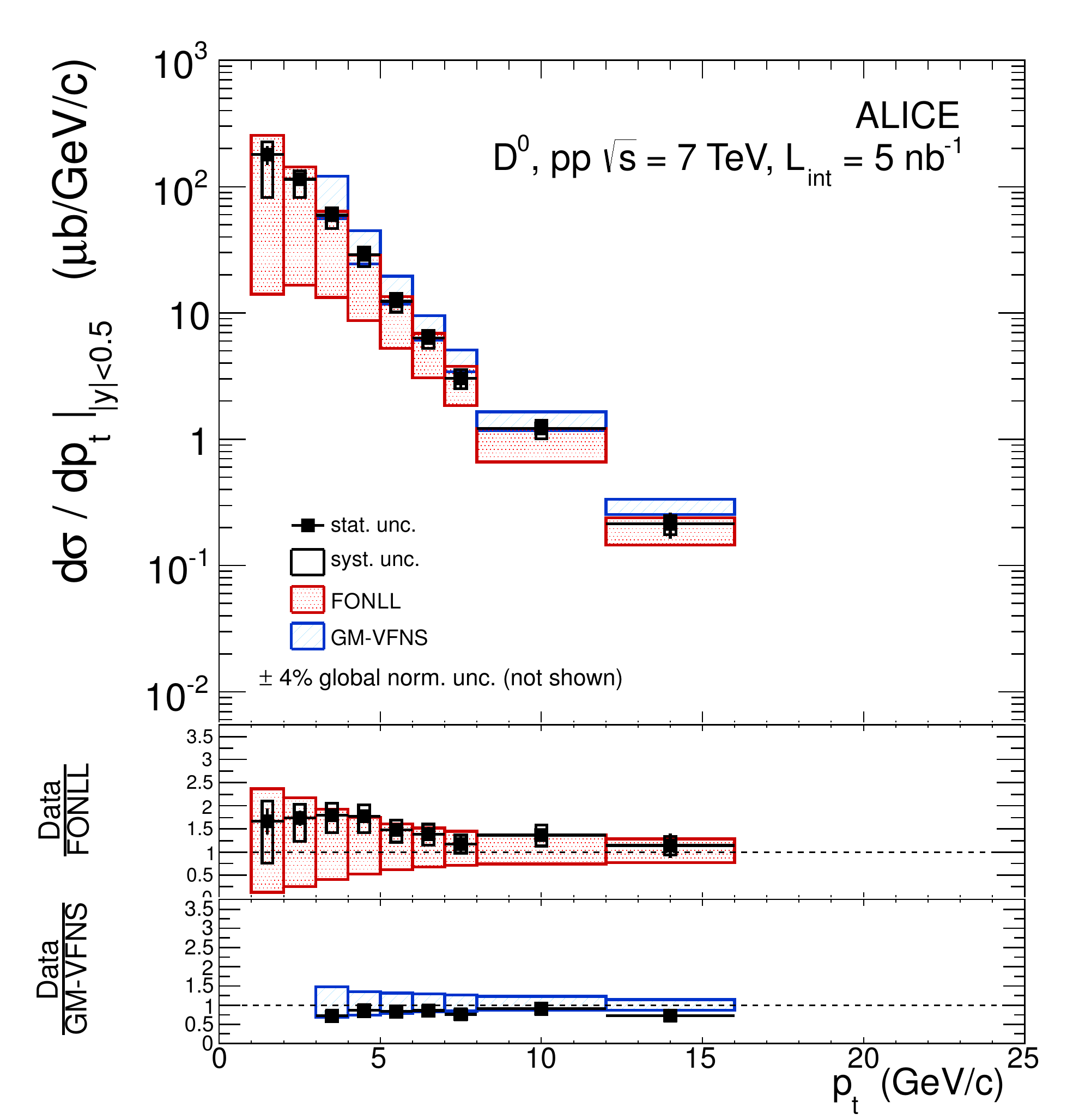} &
\hspace{-0.8cm}
\includegraphics[height=5cm]{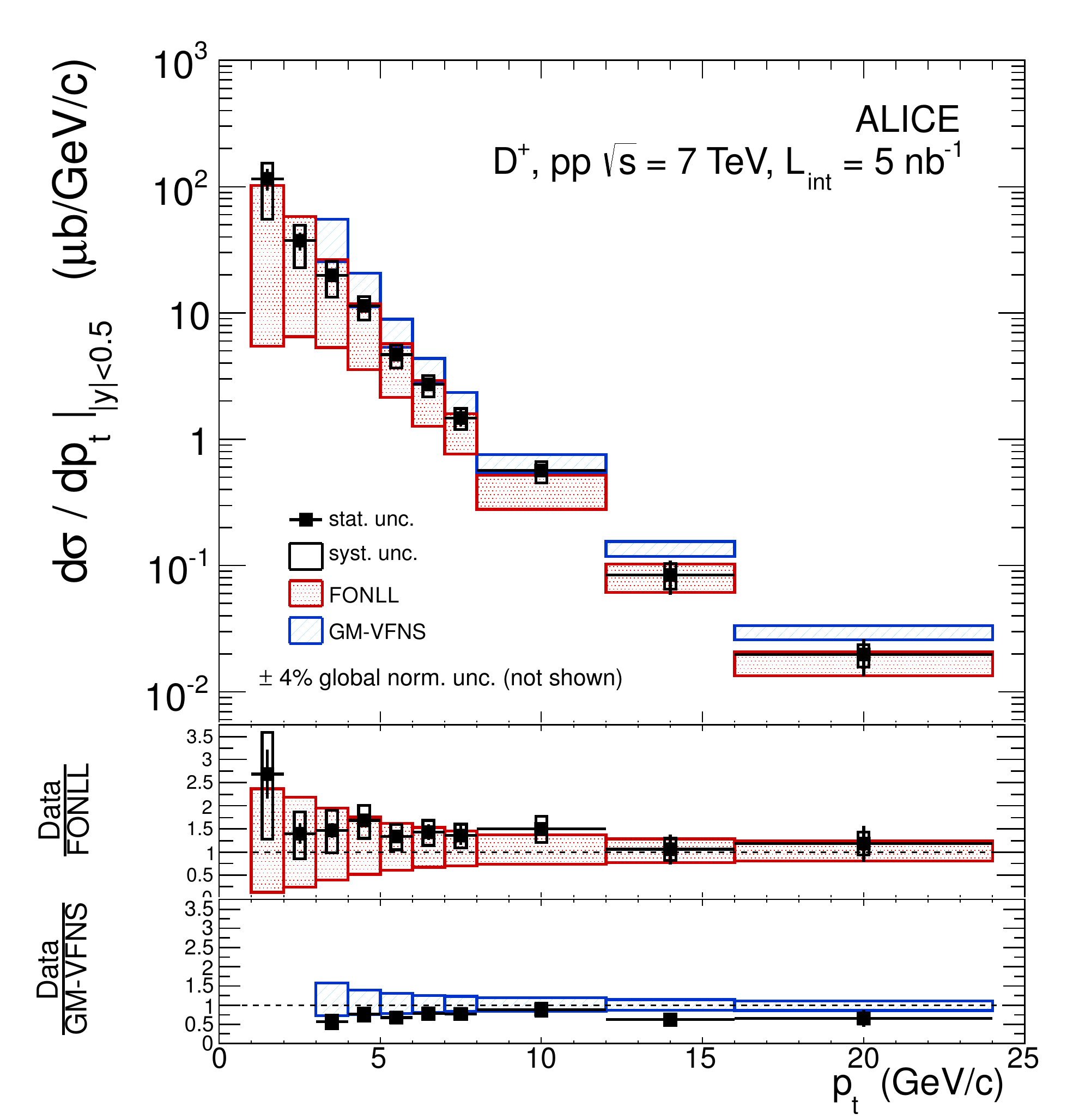} &
\hspace{-0.8cm}
\includegraphics[height=5cm]{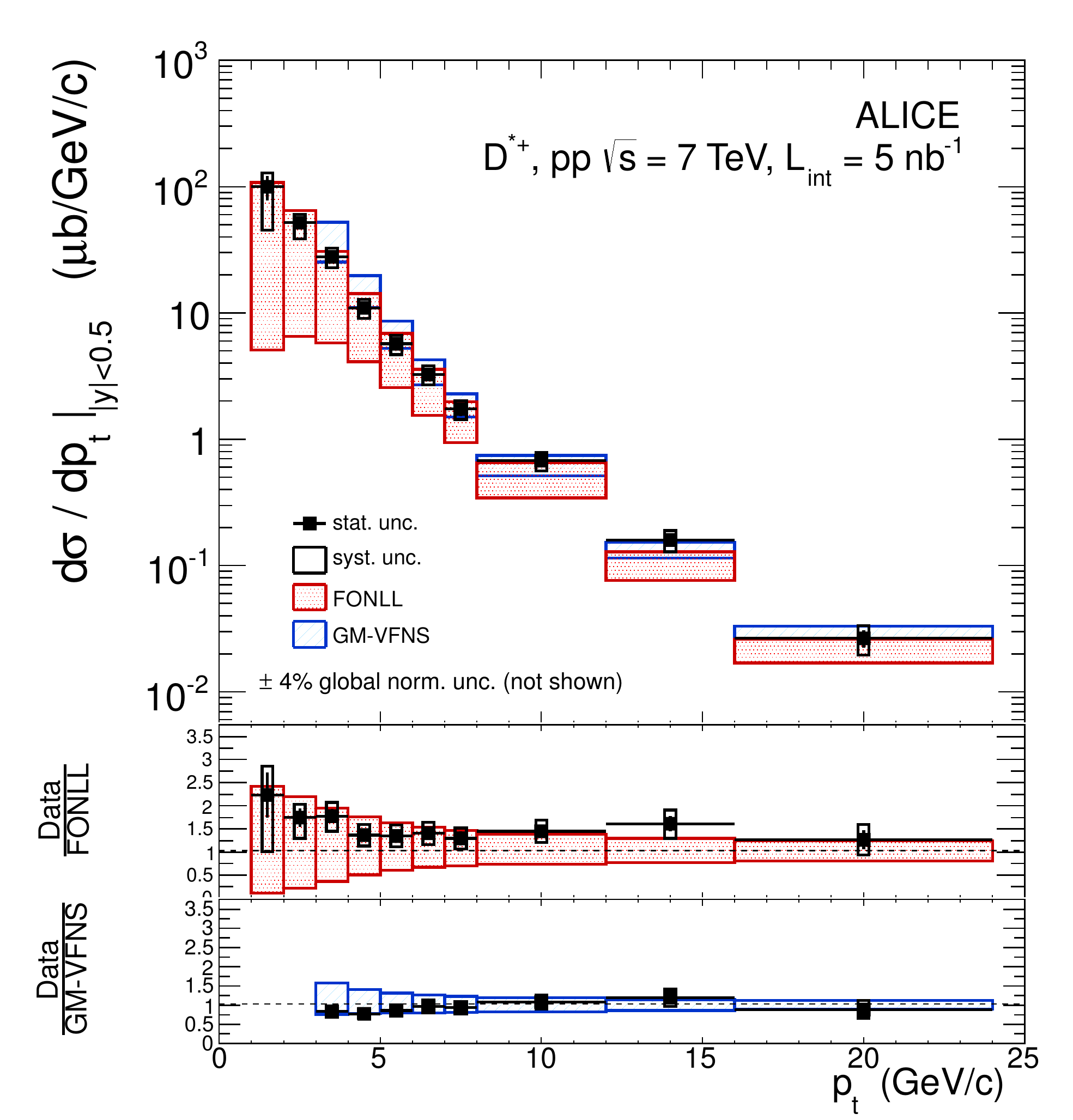} 
\end{tabular}
\end{center}
\vspace{-0.5cm}
\caption{$\Dzero$ (left), $\Dplus$ (middle), and $\Dstar$ (right) inclusive cross sections as a function of $p_t$ measured by ALICE in $\pp$ collisions at $\sqrt{s} = 7$ TeV~\cite{ALICE:2011aa}. } 
\label{figure:DcrossSection}
\end{figure}

Figure~\ref{figure:DcrossSection}~\cite{ALICE:2011aa}
shows the $\Dzero$, $\Dplus$ and $\Dstar$ $\pt$-differential inclusive cross sections in $|y| < 0.5$, 
obtained 
from $\pp$ events at $\sqrt{s} = 7$ TeV. Efficiency and acceptance corrections were applied to the 
raw data, after the B feed-down correction was estimated using FONLL predictions and applied~\cite{Cacciari}. 
In this figure (as well as in the following figures, unless specified otherwise), the error bars refer to statistical uncertainties, while boxes correspond to systematic uncertainties.
The measured cross sections are found to be well described within uncertainties by the predictions from the FONLL and GM-VFNS calculations, based on perturbative QCD~\cite{Cacciari,VFNS}. For more details on the analysis see~\cite{ALICE:2011aa}.

\begin{figure}[t!]
\begin{center}
\hspace{-0.5cm}
\includegraphics[height=6cm]{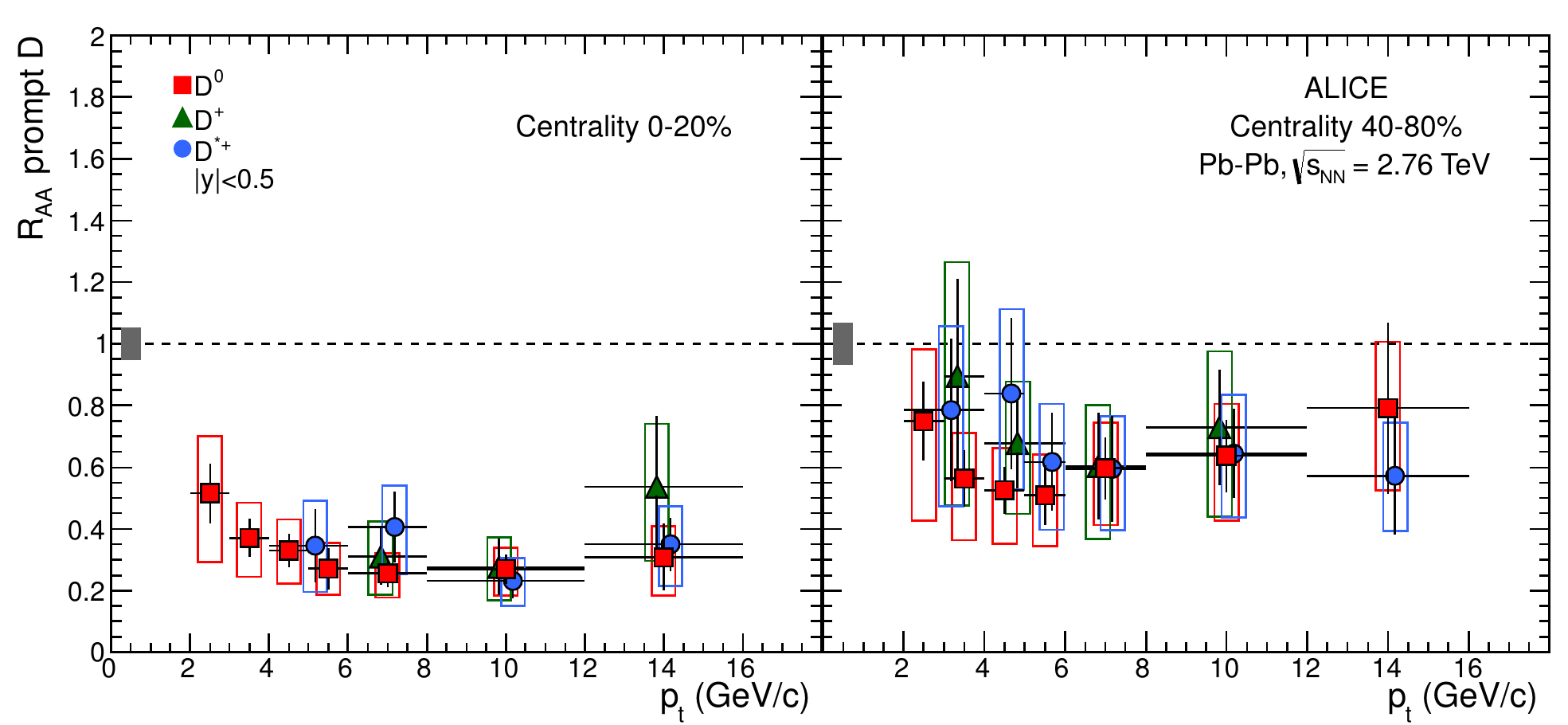} 
\end{center}
\vspace{-0.5cm}
\caption{ALICE prompt $\Dzero$, $\Dplus$, and $\Dstar$ $R_{AA}$ as a function of $\pt$ in two different centrality classes: the left panel corresponds to $0\--20\%$ (central) events, 
while the right panel shows the
results for $40\--80\%$ (semiperipheral) events~\cite{DRAA}.} 
\label{figure:DRAA}
\end{figure}

In the case of $\PbPb$ data, the D meson spectra as a function of $\pt$ and in different centrality bins\footnote{The centrality of the collision is defined in terms of percentiles of the hadronic $\PbPb$ cross section and 
determined from the distribution of the summed amplitudes in the VZERO scintillator tiles. This is fitted according to the Glauber model~\cite{Glauber} to describe the collision geometry~\cite{firstPbPbALICE}.} are obtained following the same procedure~\cite{DRAA}. These spectra are then  compared to those obtained in $\pp$
collisions. The comparison can be carried out if the $\pp$ data were taken at the same energy, and after rescaling them by $\langle T_{AA}\rangle$ 
via $R_{AA}$ (Eq.~\ref{eq:RAA}).
Due to the limited statistics of the $\pp$ data sample collected at $\sqrt{s} = 2.76$ TeV, 
which results in a limited precision and $\pt$ coverage, the $\pp$ reference spectra were obtained by using a $\sqrt{s}$-scaling of
the cross sections measured at $\sqrt{s} = 7$ TeV (see \cite{DRAA} for more details). The results for  $R_{AA}$ of prompt $\Dzero$, $\Dplus$ and $\Dstar$ mesons are shown in Fig.~\ref{figure:DRAA} for the central ($0\--20\%$ centrality class) and semi-peripheral ($40\--80\%$) events. $R_{AA}$ for the three species agree with each other in both centrality classes and over the $\pt$ ranges where they are measured (i.e. $2 \le \pt \le 16$ $\GeV/c$,  $6 \le \pt \le 16$ $\GeV/c$, $5 \le \pt \le 16$ $\GeV/c$, for 
$\Dzero$, $\Dplus$ and $\Dstar$ in the $0\--20\%$ centrality class, and $2 \le \pt \le 16$ $\GeV/c$,  $3 \le \pt \le 12$ $\GeV/c$, $2 \le \pt \le 16$ $\GeV/c$, for 
$\Dzero$, $\Dplus$ and $\Dstar$ in the $40\--80\%$ centrality class). Moreover, a clear increase in the $R_{AA}$ is visible for more peripheral collisions, 
suggesting that in more central events initial state effects are more pronounced. 
A comparison of the averaged D meson $R_{AA}$ with the charged hadrons $R_{AA}$ was carried out. Since at high $\pt$ ($\pt > 5$ $\GeV/c$) it was shown that the charged hadron $R_{AA}$ coincides with that for charged pions~\cite{Appelshauser:2011ds}, the comparison would allow to test the prediction about the colour charge and mass dependence of
energy loss, according to which heavy flavour should lose less energy than gluons, translating into $R_{AA}^{D} > R_{AA}^{charged}$. The results show quite good agreement between the heavy and light hadrons $R_{AA}$ especially at $\pt \sim 5$ $\GeV/c$. Nevertheless, there are some indications that $R_{AA}^{D}$ may be higher than $R_{AA}^{charged}$ at low $\pt$ 
(up to $\sim 30\%$ at 3 $\GeV/c$). However, this observation is not at present conclusive, due to the limited statistical precision of the current data.. 

Comparisons with different theoretical models have also been carried out. Several models
describe reasonably well both the 
charm $R_{AA}$ and the light flavour $R_{AA}$. For more details on the analysis, see~\cite{DRAA}.

Another ongoing analysis of D mesons addresses the study of elliptic flow. Details and results from this analysis can be found in~\cite{RossiHere}.

\subsection{Electrons from heavy flavour decays}
\label{sec:HFelectrons}
Heavy flavour production cross sections can be studied through the semielectronic decays of open charm and open beauty. The key tool for this analysis is the 
excellent electron PID capability of the ALICE experiment. The TPC $dE/dx$ measurements together with the TOF information enable the identification of electrons in the 
low and intermediate $\pt$ region (up to $\sim 4$ $\GeV/c$). The analysis includes  (at present only for $\pp$ data) also the TRD detector to suppress the $\pi$ background. The EMCAL detector will be also added soon, in order to extend the $\pt$ coverage of the measurements.

\begin{figure}[t!]
\begin{center}
\begin{tabular}{cc}
\includegraphics[height=7cm]{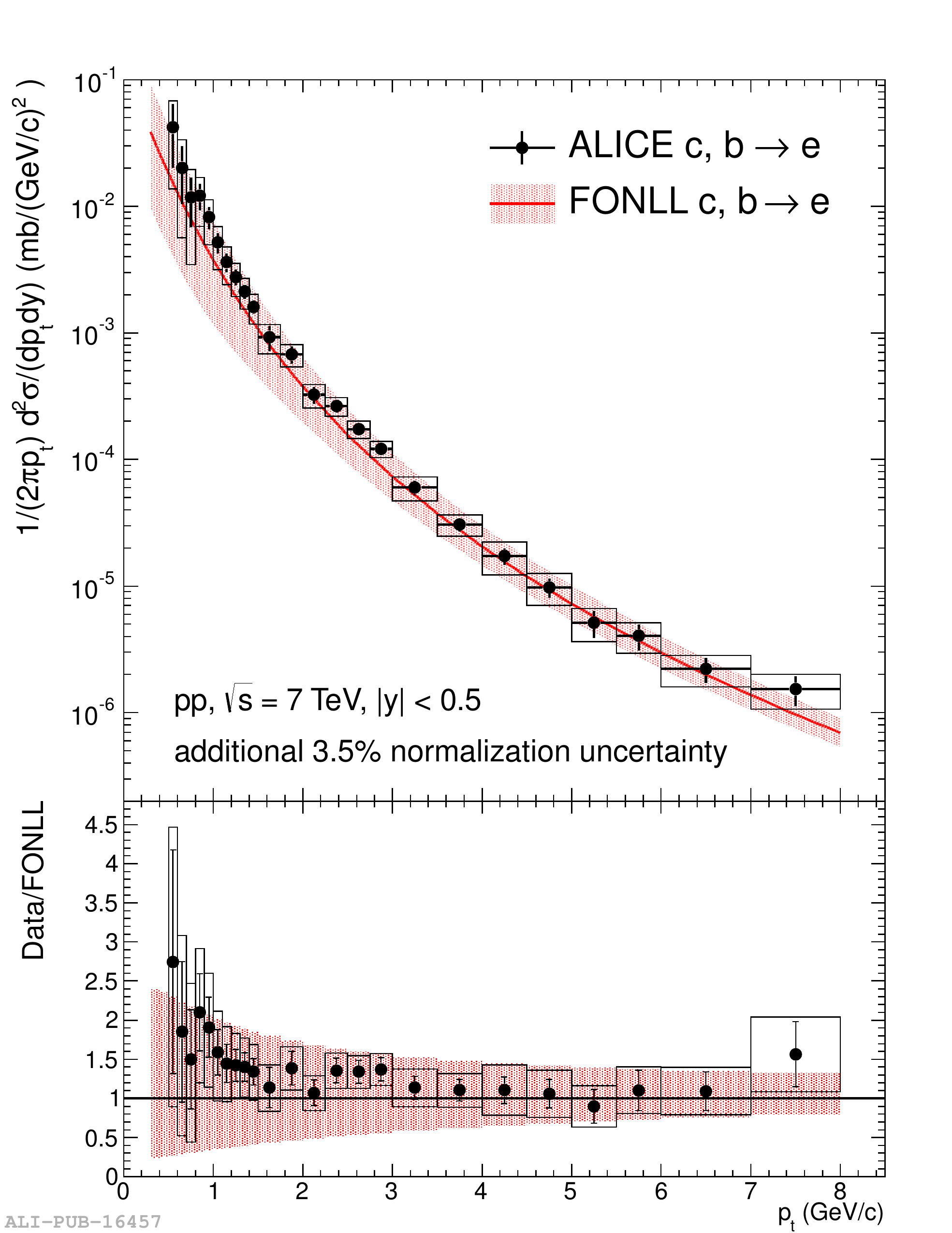} &
\hspace{-0.5cm}
 \includegraphics[height=6cm]{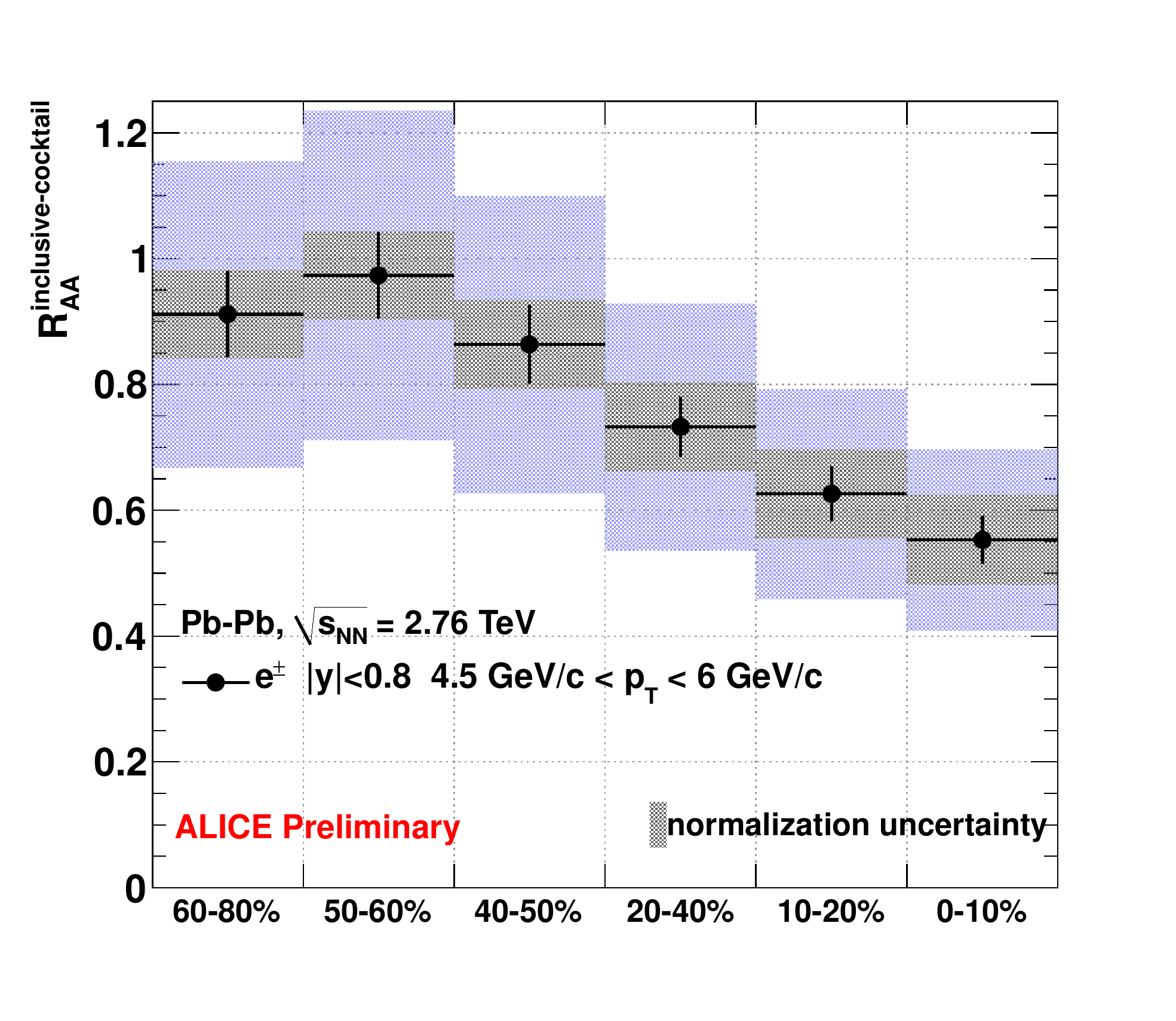}
\end{tabular}
\end{center}
\vspace{-0.5cm}
\caption{Left: $\pt$ distribution for electrons from heavy flavour measured by ALICE in $\pp$ collisions at $\sqrt{s} = 7$ TeV. Superimposed are the predictions from FONLL calculations, the ratio 
to which is presented in the lower panel. Right: $R_{AA}$ as a function of centrality for background subtracted electrons.} 
\label{figure:HFelectrons}
\end{figure}

The inclusive electron $\pt$ distributions of heavy-flavour decay electrons presented here for both $\pp$ and $\PbPb$ collision data was obtained by subtracting from the inclusive electron 
spectrum a cocktail of the measured
background sources of electrons, i.e. electrons from light hadron Dalitz decays ($\pi^0$, $\rho$, $\omega$, $\eta$), photon conversions in the material,
heavy quarkonia (J/$\psi$ and $\Upsilon$), and direct radiation.
The left panel of Fig.~\ref{figure:HFelectrons} shows the $\pt$ spectrum obtained in $\pp$ collisions at $\sqrt{s} = 7$ TeV. The result from the cocktail analysis that allows to single out electrons 
from charm and beauty decays is presented. The distribution is compared to FONLL calculations, with which good agreement can be observed.

The same analysis approach was used in the case of $\PbPb$ data at $\sqrt{s_{NN}}=2.76$ TeV. In this case, the $\pt$ distributions were obtained with the cocktail approach as a function of $\pt$ 
and centrality. It was found that in central collisions a residual background at low $\pt$ remained after the subtraction of the cocktail. This background is larger for more central collisions, possibly due to the contribution of thermal photons that are not
included in the background removal. 
For each centrality class, the $R_{AA}$ of background subtracted electrons is determined by 
taking as a reference the $\pp$ results at $\sqrt{s} =7$ TeV, scaled to $\sqrt{s} = 2.76$ TeV using FONLL 
calculations, and scaling the ratio by $\langle T_{AA} \rangle$. It was found that in peripheral collisions the value of $R_{AA}$ is compatible with one over the full $\pt$ range measured (i.e. $1.5 < \pt < 6$ $\GeV/c$), suggesting the absence of medium effects. On the contrary,
a suppression can be seen increasing with centrality in the high $\pt$ interval 
($3.5< \pt < 6$ $\GeV/c$) where the charm and beauty decay 
contribution should dominate the electron spectrum according to the comparison between the 
inclusive electron spectrum and the cocktail. 
The $R_{AA}$ measured as a function of centrality is reported in the right panel of Fig.~\ref{figure:HFelectrons}, integrating over the momentum
range $4.5 < \pt < 6$ $\GeV/c$. As one can see the $R_{AA}$ decreases from $\sim 0.9 \div 1$ to $\sim 0.55$ going from peripheral to central collisions, consistent with heavy flavour meson suppression due to partonic energy loss in the hot and dense medium created in central events. For more
details see~\cite{Raphaelle}.

\subsection{Muons from heavy flavour decays}
\label{sec:HFmuons}
Another important channel for the study of heavy flavour production is the semimuonic decay of charm and beauty. Muons coming from heavy flavour decays are reconstructed using the MUON spectrometer of ALICE. The MUON spectrometer, with both triggering and tracking chambers, 
covers the pseudorapidity region $-4 < \eta < -2.5$ and detects muons with momentum larger than 4 $\GeV/c$. The analysis strategy relies on event and track selection with subsequent subtraction of the background due to light hadrons (especially
$\pi$ and K from primary and secondary decays). The background subtraction is carried out using Monte Carlo simulations. For more details on the analysis strategy, see~\cite{Abelev:2012pi}.

\begin{figure}[t!]
\begin{center}
\includegraphics[height=6cm]{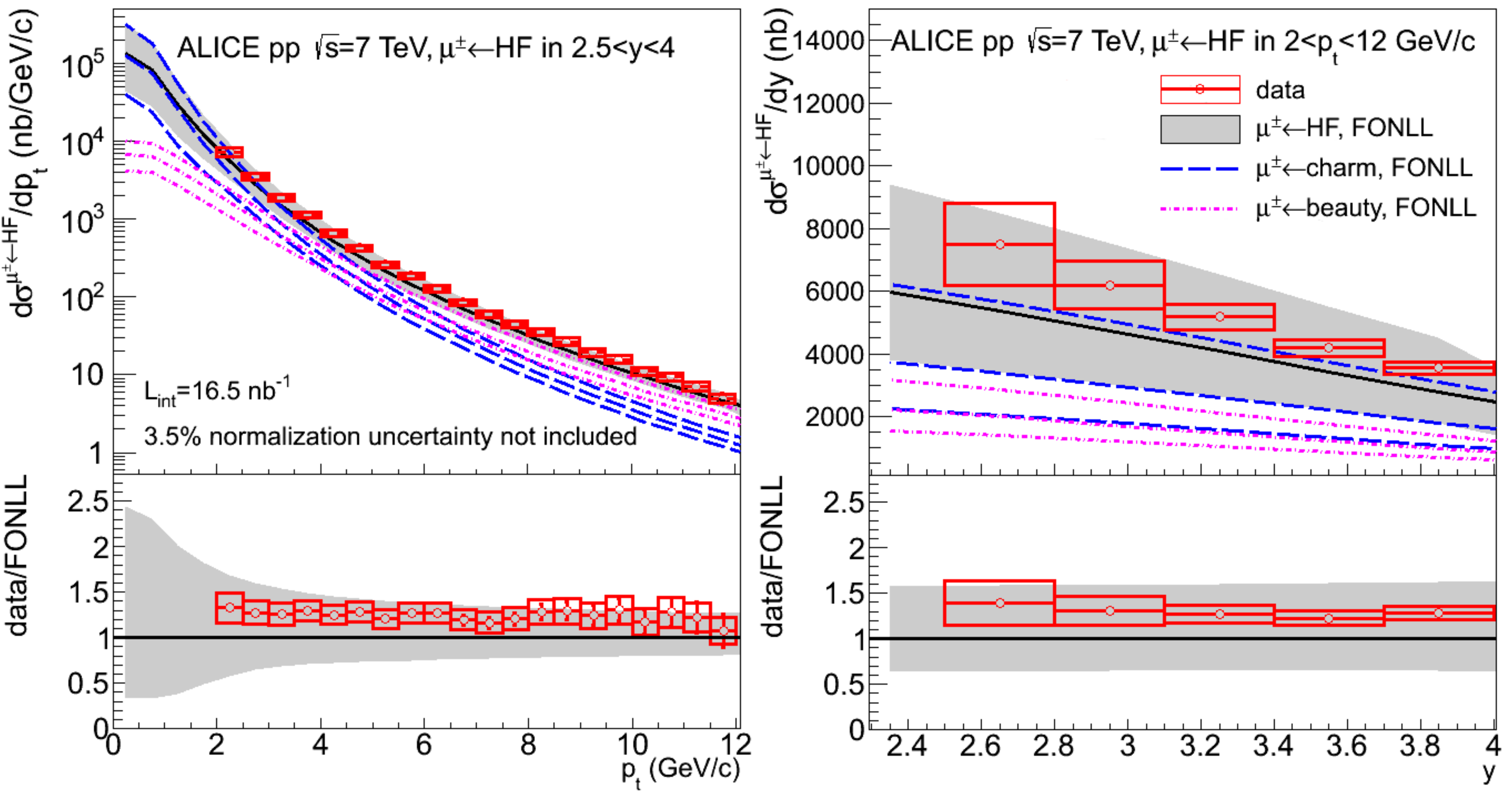} 
\end{center}
\vspace{-0.5cm}
\caption{$\pt$ (left) and $y$ (right) differential inclusive cross section for muons from heavy flavour decays measured in $\pp$ collisions at $\sqrt{s} = 7$ TeV. Also shown are the predictions from 
FONLL calculations, considering the contributions from charm and bottom both together and separately~\cite{Abelev:2012pi}.} 
\label{figure:HFmuons}
\end{figure}

Figure~\ref{figure:HFmuons} shows the $\pt$ (left panel) and the rapidity (right panel) differential cross section for the reconstructed muons after background subtraction, efficiency and acceptance
correction, obtained for $\pp$ data at $\sqrt{s} = 7$ TeV. The comparison with 
theoretical calculations from FONLL are also reported. There is in general good agreement with the predictions for both the $\pt$ and the $y$ distributions. The measured data lie above the central values of the FONLL calculation, but within the systematic uncertainties.

In order to extract the $\pt$ distributions for heavy flavour muons in $\PbPb$ collisions at $\sqrt{s_{NN}} = 2.76$ TeV, the same strategy was used. In this case, however, the background
from light hadrons was not subtracted. It is estimated to be of the order of $6-15\%$ ($2-9\%$) for $\pt > 4$ $\GeV/c$ ($\pt > 6$ $\GeV/c$) in central and peripheral collisions. This 
values were obtained from a minimum bias simulation based on the HIJING generator~\cite{Wang:1991hta} (without quenching). 

The ratio of inclusive muons in central to peripheral events is defined as:
$$
R_{CP}(\pt) = \frac{[1/\langle T_{AA} \rangle \times dN/ d \pt]_{central}}{[1/\langle T_{AA} \rangle \times dN/ d \pt]_{peripheral(60\--80\%)}}
$$

\begin{figure}[t!]
\begin{center}
\includegraphics[height=5cm]{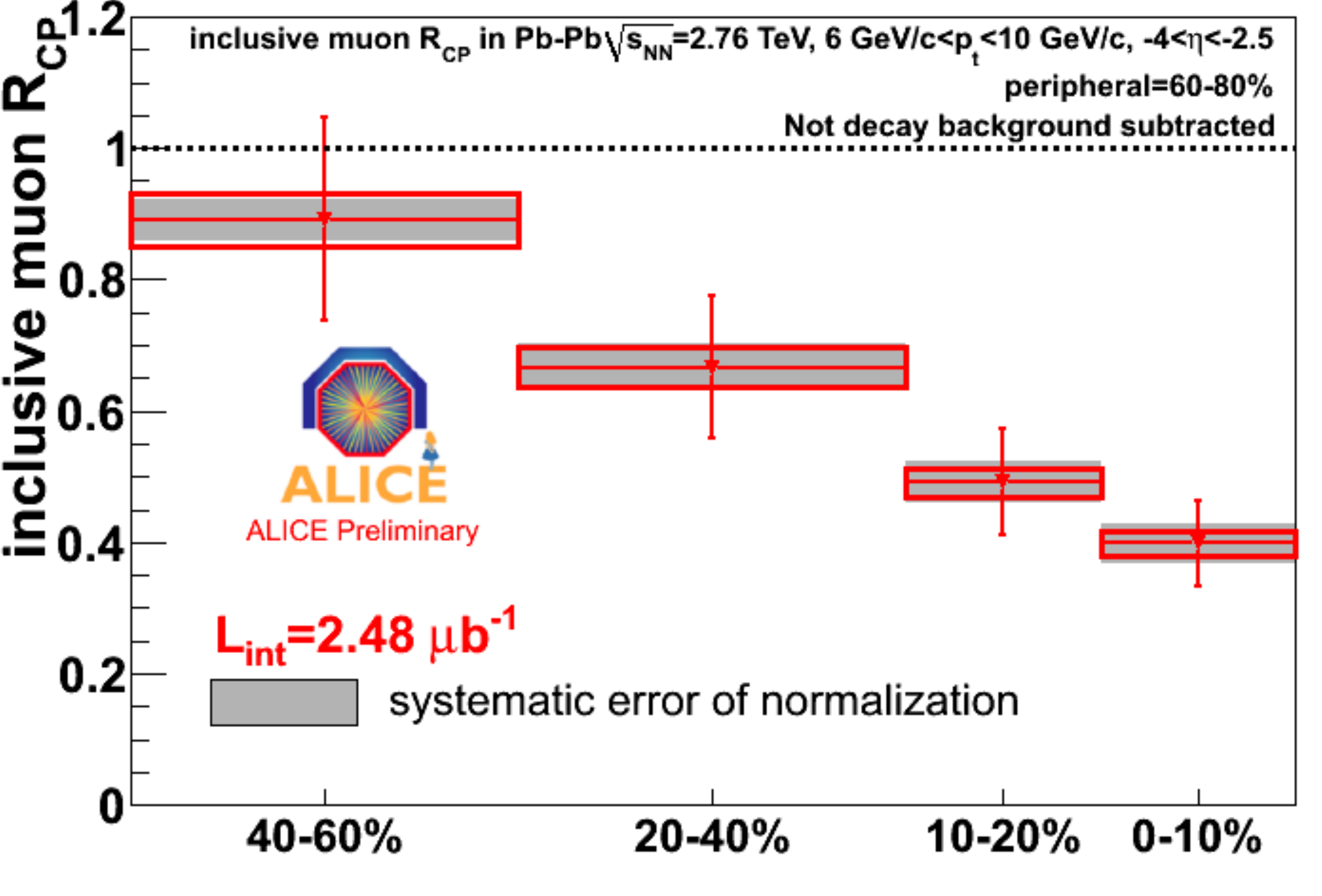} 
\end{center}
\vspace{-0.5cm}
\caption{Inclusive $R_{CP}$ for heavy flavour muons measured in $\PbPb$ collisions at $\sqrt{s_{NN}} = 2.76$ TeV. The reference centrality is the $60\--80\%$ class.} 
\label{figure:muonRCP}
\end{figure}

The Inclusive $R_{CP}$ for heavy flavour muons is shown in Fig.~\ref{figure:muonRCP}. Here only muons with $\pt > 6$ $\GeV/c$ are considered in order to keep the background under control. As one can see, the suppression with respect to peripheral 
collisions becomes larger as centrality increases, from $\sim 0.9$ for the most peripheral to $\sim 0.4$ for the most central events. 
This could be interpreted as a consequence of the formation of a hot and dense medium in central events. For more details see also~\cite{Lopez}.

\section{Quarkonia}
\label{sec:Jpsi}
In this section, some of the J/$\psi$ measurements performed with the ALICE detector will be described, both from the analysis of the $\pp$ data at $\sqrt{s} = 7$ TeV and 2.76 TeV, and from the 
$\PbPb$ data sample collected at $\sqrt{s_{NN}} = 2.76$ TeV. 
The nuclear modification factor $R_{AA}$ will be presented. 
The measurements are carried out for the dielectron decay channel at midrapidity ($|y| < 0.9$), and the dimuon channel at forward rapidity ($-4<y<-2.5$). For more details about
the ALICE J/$\psi$ results, refer to~\cite{Aamodt:2011gj,Abelev:2011md,Abelev:2012rv,Abelev:2012rz,Abelev:2012kr}.

\begin{figure}[t!]
\begin{center}
\begin{tabular}{cc}
\includegraphics[height=6cm]{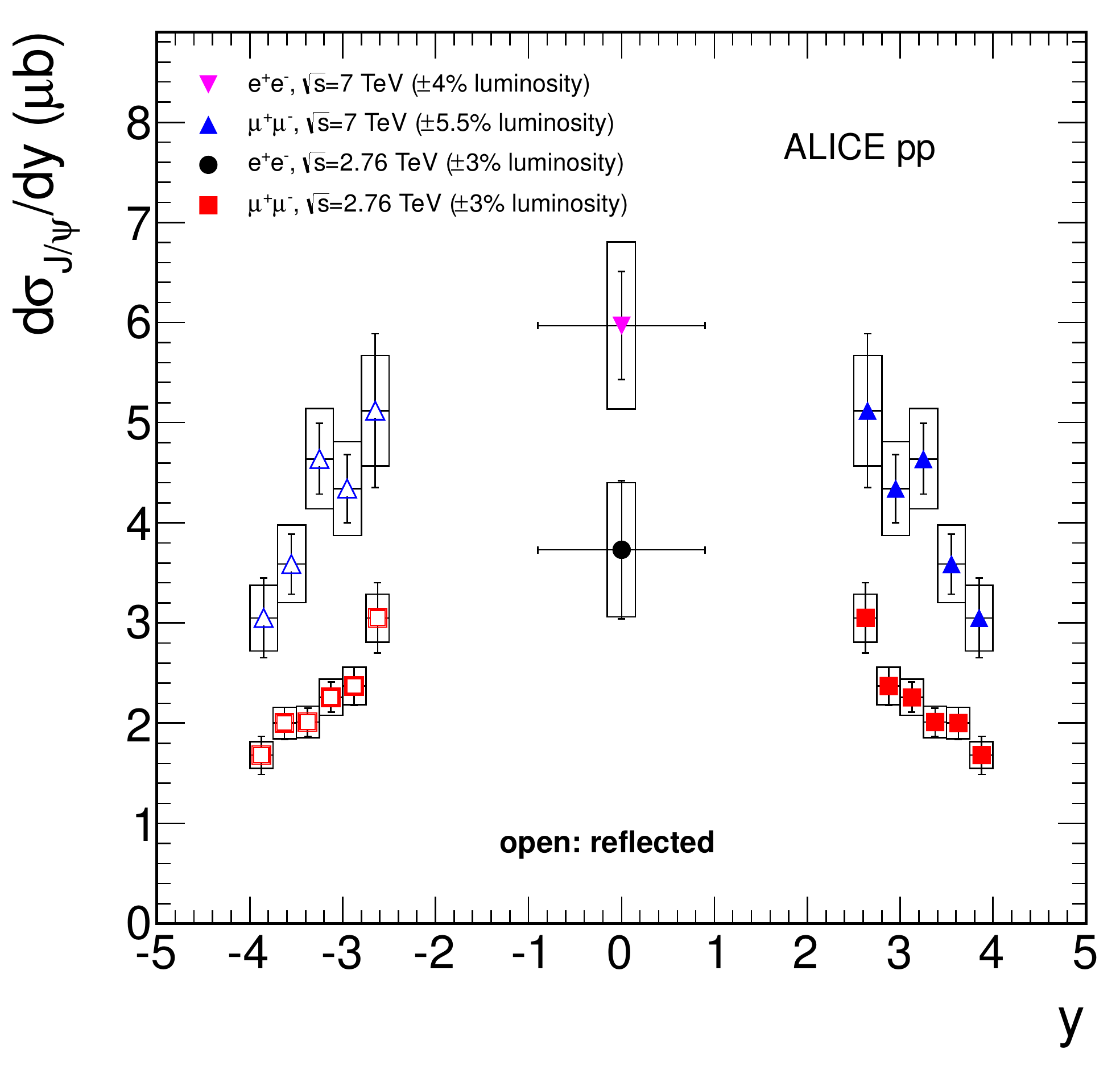}
\includegraphics[height=6cm]{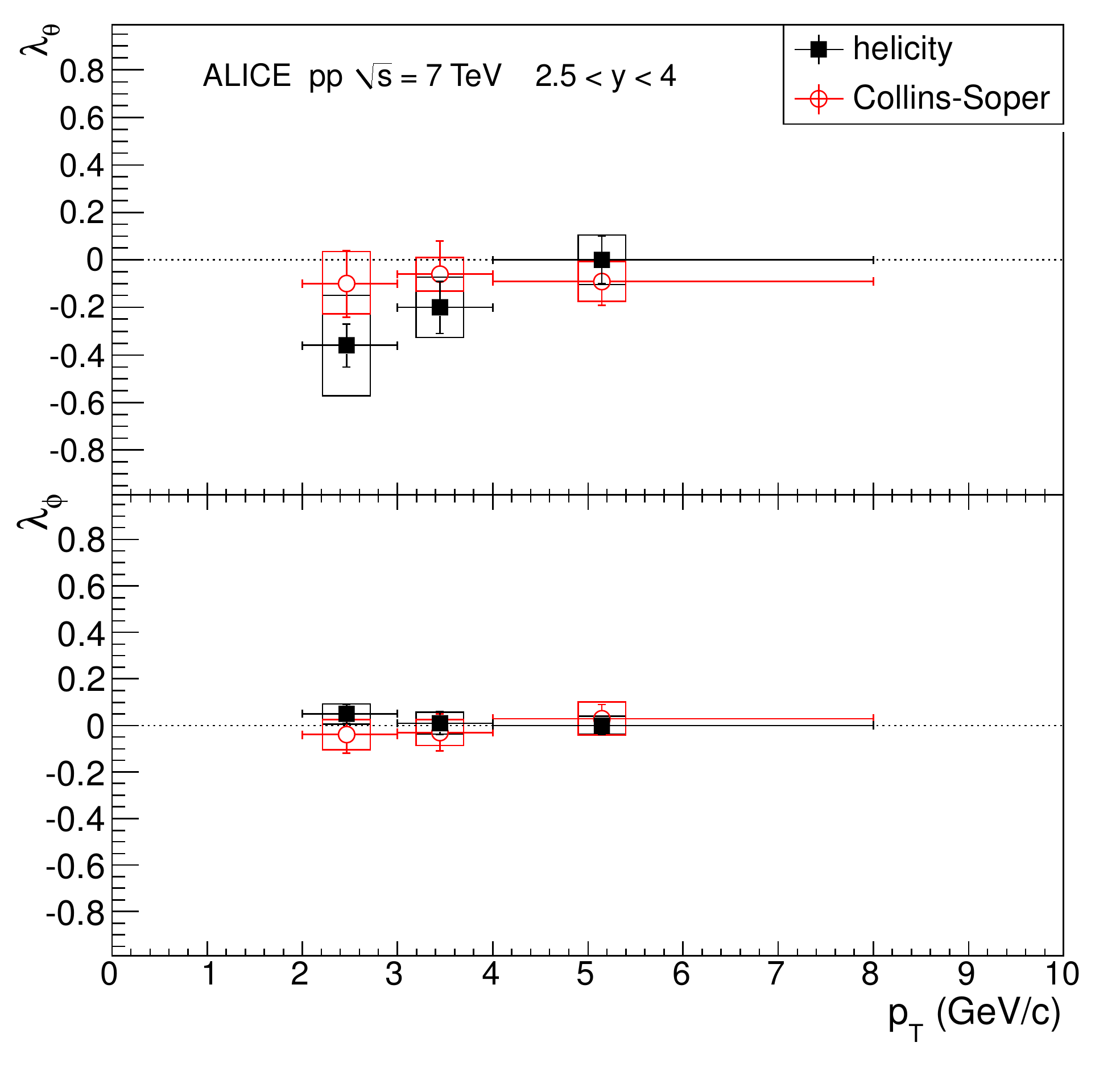}
\end{tabular}
\end{center}
\vspace{-0.5cm}
\caption{Left: inclusive $\pt$-integrated J/$\psi$ cross section measured in $\pp$ collisions at 7 TeV. The results include the measurement at midrapidity and at forward rapidity. 
The measurement at $\sqrt{s} = 2.76$ are shown~\cite{Abelev:2012kr}. Right: $\lambda_{\theta}$ and $\lambda_{\phi}$ J/$\psi$ polarization parameters as a function of $\pt$ obtained in $\pp$ collisions at $\sqrt{s} = 7$ TeV~\cite{Abelev:2011md}.} 
\label{figure:Jpsi}
\end{figure}

The J/$\psi$ analysis at central rapidity relies on the PID information provided
by the TPC detector. In future the analysis will include the TRD, TOF and EMCAL to improve the electron PID and the triggering capabilities. 
The signal extraction was performed via bin counting in the invariant mass range $2.92 < M_{inv} < 3.16$ GeV/$c^2$, with background removal performed by subtracting from 
the opposite-sign electron pairs' invariant mass distribution the corresponding like-sign distribution. The inclusive $\pt$-integrated cross section measured in $\pp$ collisions at 7 TeV
is shown in the left panel of Fig.~\ref{figure:Jpsi}, together with the results for $\pp$ collisions at $\sqrt{s} = 2.76$ TeV (see~\cite{Abelev:2012kr} for more details).

In the same figure, the results from the J/$\psi$ analysis at forward rapidity, i.e. $-4 < y<-2.5$,
are included. In this rapidity interval, the measurement was carried out using the MUON spectrometer. The signal was extracted by fitting the data
with a Crystal-Ball function, a technique used also for the analysis of $\PbPb$ data. 

The measurement of polarization in $\pp$ collisions at the LHC offers a way to further test of QCD theoretical calculations for the production mechanism of quarkonia. This may discriminate between the various QCD approaches that are not at present able to describe all quarkonia observables consistently. The right panel of Fig.~\ref{figure:Jpsi}
shows the polarization parameters $\lambda_{\theta}$ and $\lambda_{\phi}$ obtained from the polar and azimuthal angular distributions of the J/$\psi$ decay
muons measured at forward rapidity 
($-4 < y < -2.5$) in the helicity and Collins-Soper reference frames~\cite{Abelev:2011md}. As one can see, both $\lambda_{\theta}$ and $\lambda_{\phi}$ are compatible with zero in both reference frames, even if a decrease of $\lambda_{\theta}$ to $\sim -0.4$
in the lower $\pt$ range is visible in the helicity reference frame. 

\begin{figure}[t!]
\begin{center}
\includegraphics[height=6cm]{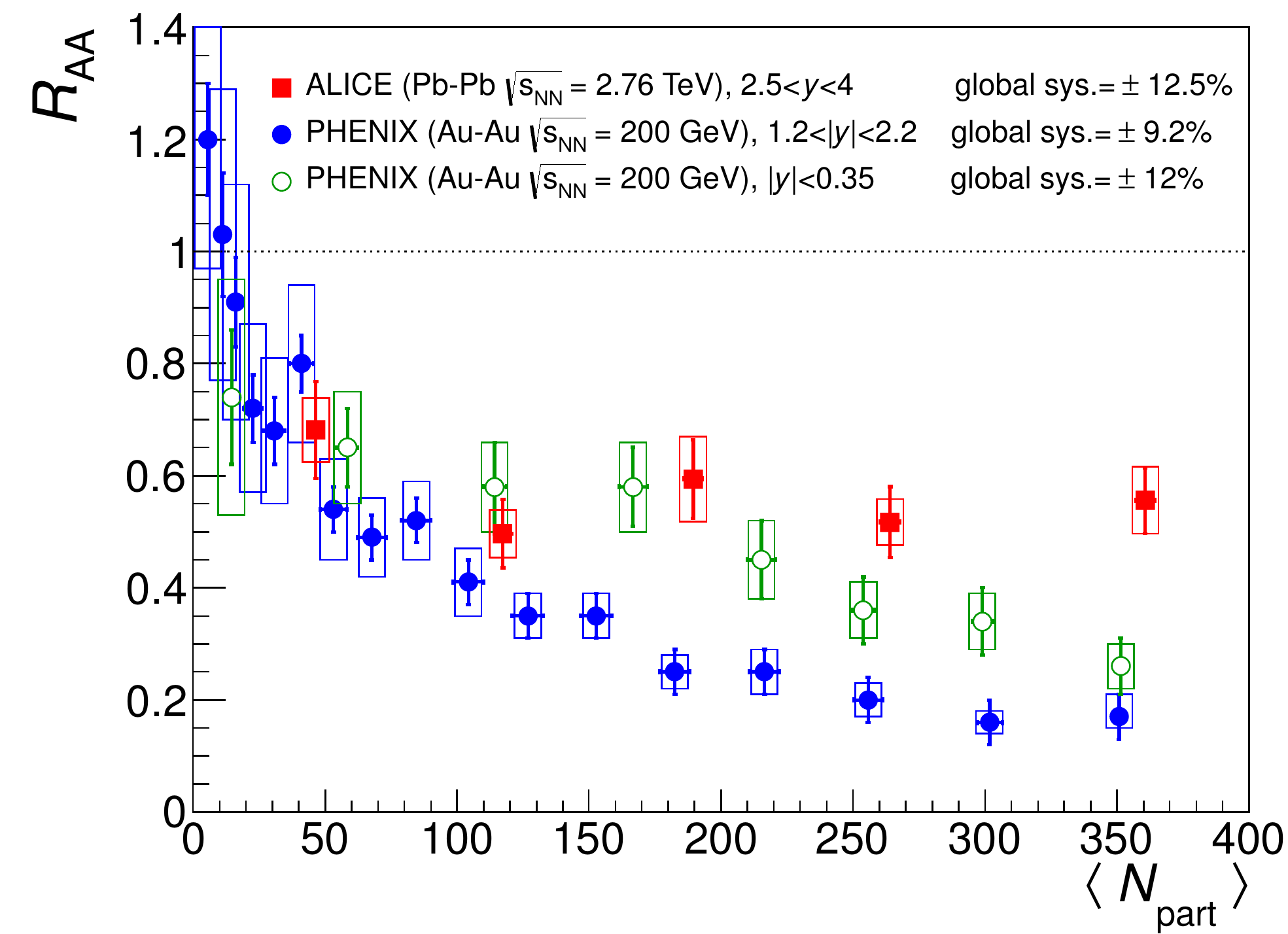}
\end{center}
\vspace{-0.5cm}
\caption{J/$\psi$ $R_{AA}$ as a function of the number of participating nucleons obtained by ALICE 
in $\PbPb$ collisions at $\sqrt{s_{NN}} = 2.76$ TeV. The measurements by PHENIX in Au\--Au collisions at 
$\sqrt{s_{NN}} = 0.2$ TeV are also shown~\cite{Abelev:2012rv}.} 
\label{figure:JpsiRAA}
\end{figure}

Combining the information on the J/$\psi$ yield in $\PbPb$ collisions at $\sqrt{s_{NN}} = 2.76$ TeV in a given centrality class with that from $\pp$ collisions at the same centre-of-mass energy, it was possible to 
obtain the $R_{AA}$ for J/$\psi$ in the forward region~\cite{Abelev:2012rv}. 
This is shown in figure~\ref{figure:JpsiRAA}, where the J/$\psi$ $R_{AA}$ is drawn as a function of the number of participating nucleons, used
as the parameter to describe the centrality of the collisions (the higher $\langle N_{part} \rangle$, the more central the event is). As one can see, the $R_{AA}$ shows little dependence on centrality
staying at the level of $\sim 0.7 \div 0.5$. Moreover,
the comparison with the results from PHENIX in Au\--Au collisions at $\sqrt{s_{NN}} = 0.2$ TeV~\cite{Adare:2006ns,Adare:2011yf,Adler:2004zn}, which are also reported in the figure, indicates a smaller suppression at LHC than at RHIC in central collisions. Models based on statistical hadronization, or including J/$\psi$ regeneration from charm quarks in the QGP phase can describe the data quite
well, as discussed in~\cite{Abelev:2012rv}.

\hspace{-1cm}
\section{Conclusions}
\label{sec:conclusions}
The main heavy flavour results by the ALICE experiment at the LHC have been presented. The importance of the analysis of $\pp$ data has been shown in terms of QCD predictions benchmark, 
and as a reference for the $\PbPb$ measurements. The $R_{AA}$ of open heavy flavour exhibits a suppression of $\sim 4$ for D mesons in $0\--20\%$ centrality class events. 
The D meson results show an increasing suppression with centrality, seen also for electrons and muons from heavy flavour decays, which indicate that in central collisions a dense and strongly interacting medium seems to be formed.  
The measurement of $R_{AA}$ of J/$\psi$ at LHC shows a different behavior as a function of centrality compared to measurements at RHIC,  which could be an indication of (re)generation of 
J/$\psi$ in  the QGP (or at the phase boundary).  

The forthcoming p$\--$Pb collisions foreseen for November 2012 will offer the opportunity to better understand the results of the ALICE open heavy flavour and charmonium results. They will 
provide key measurements to discriminate between initial and final state effects, such as shadowing, and to evaluate the degree of thermalization of open charm.

\begin{spacing}{0}

\end{spacing}
\end{document}